# ScamGPT-J: Inside the Scammer's Mind, A Generative AI-Based Approach Toward Combating Messaging Scams

Xue Wen Tan
*National University Of Singapore*, xuewen@u.nus.edu

Kenneth See
*National University of Singapore*, see.k@u.nus.edu

Stanley Kok
*National University of Singapore*, skok@comp.nus.edu.sg

# ScamGPT-J: Inside the Scammer's Mind
# A Generative AI-Based Approach Toward Combating Messaging Scam

*Completed Research Paper*

**Xue Wen Tan**[*]
Asian Institute of Digital Finance
National University of Singapore
xuewen@u.nus.edu

**Kenneth See**[*]
Asian Institute of Digital Finance
National University of Singapore
see.k@u.nus.edu

**Stanley Kok**
Department of Information Systems and Analytics
National University of Singapore
skok@comp.nus.edu.sg

## Abstract

*The increase in global cellphone usage has led to a spike in instant messaging scams, causing extensive socio-economic damage with yearly losses exceeding half a trillion US dollars. These scams pose a challenge to the integrity of justice systems worldwide due to their international nature, which complicates legal action. Scams often exploit emotional vulnerabilities, making detection difficult for many. To address this, we introduce ScamGPT-J, a large language model that replicates scammer tactics. Unlike traditional methods that simply detect and block scammers, ScamGPT-J helps users recognize scam interactions by simulating scammer responses in real-time. If a user receives a message that closely matches a ScamGPT-J simulated response, it signals a potential scam, thus helping users identify and avoid scams more effectively. The model's effectiveness is evaluated through technical congruence with scam dialogues and user engagement. Our results show that ScamGPT-J can significantly aid in protecting against messaging scams.*

## Introduction

The frequency of online scams has escalated with the proliferation of the internet, leveraging the convenience and anonymity it offers. In recent years, there has been a significant rise in scams conducted via instant messaging platforms, with an estimated loss of US$1.026 trillion in 2023 alone, approximately 58% of which is attributed to messaging platform scams (Abraham et al., 2023). It has also been observed that online scams are becoming increasingly sophisticated, with scam syndicates adopting state-of-the-art technology (Cross, 2022) and capitalizing on current trends (Ma & McKinnon, 2021) to increase the efficacy and velocity of their scam efforts. This situation, albeit regrettable, poses a pivotal challenge and concurrently offers a significant opportunity for the field of Human-Computer Interaction (HCI) to develop and implement technological countermeasures.

Current HCI literature on scam prevention highlights several critical points that reinforce this concern. Firstly, simply predicting whether a message is a scam is insufficient, as most attacks cannot be automatically detected with certainty (Zhuo et al., 2023). Secondly, while awareness and education are essential for empowering users to identify scams and avoid falling victim to them  (Reinheimer et al., 2020; Wash, 2020), their efficacy remains uncertain, with little empirical evidence supporting their best practices (Shillair et al., 2022). Additionally, large-scale awareness campaigns, such as those implemented through public policies or community outreach programs, differ from individual education efforts. Although these campaigns aim to inform the general public about scam prevention on a broader scale, their impact is often limited, as individuals may not fully comprehend the advice provided or may lack the motivation to apply it effectively (Bada et al., 2015). Furthermore, this challenge is compounded by the tendency of individuals to misjudge their own vulnerability to scams (Downs et al., 2006). Lastly, Aneke et al. (2021) propose an approach to scam prevention that assists users in recognizing phishing scams by generating explanation messages whenever a phishing attempt is detected. While this method aims to improve user awareness and decision-making, its practical usefulness and relevance may be limited. The primary concern is that the interpretations of these explanation messages can vary significantly among different users and experts, potentially leading to inaccuracies.

---

[*] Both authors contributed equally as first co-authors.

The ongoing exploitation of individuals through progressively sophisticated scamming techniques highlights the urgent need for enhanced defensive measures against such frauds. Recent studies highlight a notable shift in the psychological state of scam victims (Dove, 2020), a dimension insufficiently addressed in current HCI frameworks. Existing HCI methods fall short when it comes to helping victims acknowledge their engagement in scams. Scammers skillfully apply strategies that disrupt the victims' capacity for rational thought, causing affected individuals to often find themselves emotionally charged (Dove, 2020) or caught in cognitive biases like the sunk cost fallacy (Van Dijk & Zeelenberg, 2003). In such mental states, logical reasoning is diminished (Jung et al., 2014), making straightforward advice less effective. For example, there have been instances where bank employees' efforts to discourage clients from transferring money to probable scammers were unsuccessful (Tan & Devaraj, 2023). To address this challenge, we introduce ScamGPT-J, a novel approach designed to help victims self-identify scam scenarios.

ScamGPT-J at its core is a fine-tuned generative Large Language Model (LLM) that uses historical dialogue patterns to replicate scammer-like interactions, presenting a significant advancement in the efforts to thwart scam operations. We anchor our approach around the psychological concepts of *Heuristic and Systematic Information Processing* (Chaiken, 1980) and the *Anchoring Effect* (Furnham & Boo, 2011), which is discussed in detail in Section 3. The underlying hypothesis of our approach is that if an analogical argument is made to allow a victim to observe the high degree of similarity between the responses generated by ScamGPT-J and an actual scammer's communication, it could significantly enhance his recognition and acceptance of being a scam victim.

Our paper's key contributions are as follows:

- We develop a novel interaction framework that encompasses scammers, victims, and ScamGPT-J. This framework harnesses the capabilities of LLMs to generate authentic scammer-like conversations, thereby empowering users to independently discern potential scams. This approach significantly differs from traditional binary methods of scam identification, offering a more user-centric solution.
- We propose ScamGPT-J, a new LLM specifically fine-tuned to simulate scam responses. We also address the challenge posed by the dearth of scam conversations that can serve as training data and build a synthetic instant messaging scam conversation dataset that can be utilized for future work in this line of research.
- We implement a novel evaluation survey architecture for user experience assessment, enabling seamless interaction with ScamGPT-J. We carefully design a controlled experimental environment to minimize biased feedback from users, ensuring the reliability and accuracy of our experiment results.
- To our knowledge, this is the first study in HCI literature to apply an LLM in the context of scam detection and prevention.

# Related Literature

## *Scam Detection and Prevention*

While spam detection in emails has evolved significantly, with modern solutions achieving over 90% accuracy (Dedeturk & Akay, 2020; Saidani et al., 2020), detecting scams, especially within the context of instant messaging, poses unique challenges. One major hurdle is the lack of quality datasets, which hampers the development of effective detection methods (Shafi'I et al., 2017). Spam detection similarly encountered this challenge, leading to the development of resources like the NUS SMS Corpus (Chen & Kan, 2013) and the UCI SMS Spam Collection (Almeida et al., 2011), which have facilitated preliminary applications of machine learning techniques in distinguishing spam text messages (Kumar & Gupta, 2024; Wijaya et al., 2023). However, it is paramount to recognize the nuances of separating spam from scam messages, where the latter are designed with malicious intent. Spam messages are generally unsolicited communications that, while annoying, are relatively harmless. In contrast, scam messages are designed to manipulate victims into actions that result in financial loss. Such losses can have profound and far-reaching consequences. Beyond the immediate depletion of resources, financial loss can lead to significant emotional and psychological stress, straining relationships and even causing family breakdowns (Kassem, 2023). In extreme cases, these stresses can trigger mental health issues such as anxiety, depression, or suicidal thoughts (Whitty & Buchanan, 2016). Scam strategies continually evolve as perpetrators adapt to new trends and vulnerabilities (Zhang et al., 2022), often employing sophisticated social engineering techniques. The scammers' adaptability leads to extensive socio-economic damage, with global financial losses from scams now exceeding half a trillion US dollars annually (Abraham et al., 2023). Accurately predicting whether a text conversation is a scam is further complicated by the prevalent use of abbreviations, emoticons, and misspellings in instant messaging, which hinders effective detection (Mishra & Soni, 2023). While Mishra and Soni (2023) propose a model for phishing SMS detection, the broader spectrum of scam messages remains largely unexplored, highlighting a significant gap in current research and suggesting a need for contributions towards a scam message dataset.

Furthermore, many of the existing detection literature often overlook the context of conversations. Scammers' messages might seem innocuous in isolation but reveal their deceptive nature when viewed as part of a conversation. The importance of analyzing entire conversations to detect and understand scams is highlighted by Edwards et al. (2017). Derakhshan et al. (2021) demonstrate that telephone-based scams can be identified with a high degree of accuracy using scam signatures, which are defined as a set of text phrases that collectively fulfill the goal of the scammer. Scam signatures can either be formulated manually through known scam characteristics or identified through clustering of a series of known scam conversations.

Beyond detection, the literature also explores methods to prevent scams. Martin et al. (2018) conduct a field study showing that individuals are more vulnerable to targeted phishing attacks compared to general ones. They utilize equal-variance signal detection theory to model susceptibility to phishing scams, suggesting the need for differentiated prevention techniques based on individual risk profiles. Another countermeasure proposed is the use of social engineering against scammers (Canham & Tuthill, 2022). Edwards et al. (2017) explore this approach, highlighting the trend of "scambaiting", where individuals engage with scammers to waste their resources.

Unlike the existing literature on scam detection that has mostly focused on classifying texts in isolation, this paper accounts for conversational context in our proposed solution. Furthermore, while current methods of scam prevention focus on prescriptive approaches (i.e., telling a user if a message is a scam or not), our proposed solution brings users to the forefront by empowering them to perform scam detection themselves. Our ScamGPT-J model addresses these challenges by mimicking the conversational style of scammers. Through training on entire scam-related conversations, the model has learned to "think" like a scammer, anticipating what to say next, to set up the scam. This generative approach contrasts with existing methods that rely on keywords or isolated sentences to predict scams. Instead, our model is capable of generating complete conversations, having been trained to generate both the user's and the scammer's dialogue. However, we have constrained the model to only show the scammer's next response. This allows the model to dynamically adapt its replies based on how the conversation unfolds, as it uses the historical conversation to generate the next few lines of dialogue, making it more effective at capturing the dynamics of scam interactions.

### *Employing LLMs to Guide Human Behavior*

Fogg (1998) suggests that computers can function as persuasive tools, influencing individual behavior. The deployment of such tools is not without merit, as they are capable of shaping desirable behavior. However, this influence raises ethical considerations that must not be overlooked. Acemoglu et al. (2024) formalize Fogg's (1998) theory in a model that shows how artificial intelligence (AI) tools could offer consumer benefits under limited usage but can result in behavior manipulation when abused.

An emerging perspective in the field of HCI concerning AI's persuasive abilities is the study of reflection theory. Abdel-Karim et al. (2023) explore how AI-based systems can encourage reflective thinking among medical practitioners. Their findings suggest that when an AI system presents information that challenges a practitioner's existing beliefs and sustains engagement over time, it can create deep cognitive dissonance, leading to reflection rather than defensiveness. They introduce a *Machine-Induced Reflection Model*, which demonstrates how AI can guide users through a reflective process, ultimately helping them make more confident and thoughtful decisions, thereby enhancing their professional effectiveness. Their work contributes to the "massive rethinking" of Information Systems theory (Teodorescu et al., 2021) by highlighting the potential of machine-induced reflections to foster a more contemplative approach to work. They also observe that while decision support and recommender systems are typically designed to aid decision-making and persuasion, they can sometimes introduce biases that undermine decision quality, such as *automation bias* (Goddard et al., 2012). In contrast, AI systems explicitly designed to promote reflection offer a promising alternative. Although their study focuses on the healthcare sector, their theoretical framework can be extended to scenarios such as scam detection, where individuals must determine if they are interacting with a scammer. Our system could similarly prompt users to "rethink" their decisions when they are at risk of falling for a scam, thereby reducing the likelihood of becoming victims and increasing their confidence in their judgment.

In a similar vein, LLMs can be conceptualized as an extension of these persuasive technologies, guiding individual actions through sophisticated language processing capabilities. Despite their inability to mimic the full spectrum of human reasoning, the potential of LLMs is noteworthy in their capacity to accurately discern sentiment in text (Tabone & de Winter, 2023). This capability enables the generation of contextually appropriate responses, making them powerful tools in conversational and behavioral guidance contexts. While originally designed for general use, LLMs can be tailored for specific applications (Demszky et al., 2023). It is important to note that off-the-shelf LLMs may fall short in specialized scenarios, necessitating further training via methods such as fine-tuning or prompt-tuning.

The capacity of LLMs to influence societal behavior has been exploited for both beneficial and malevolent purposes. Documented instances reveal the use of actors leveraging LLMs to produce human-like content on platforms like Twitter (now known as *X*), and to orchestrate campaigns for questionable websites (Yang & Menczer, 2024). Additionally, there have been instances where LLMs have played a key role in generating misinformation, with state-sponsored entities utilizing them for disinformation campaigns to advance national agendas (Ezzeddine et al., 2023).

Despite these nefarious applications, LLMs also hold the potential for positive impact. Chen and Shu (2024) recognize the detrimental effects of LLM-generated misinformation but argue that these models can act as a double-edged sword. They propose that LLMs can be employed to counteract misinformation by facilitating rapid detection of false information. Ongoing research highlights the growing emphasis on using LLMs for fact verification, offering individuals timely and accurate feedback on the information they encounter (Cheung & Lam, 2023; Zhang & Gao, 2023).

In line with the theories presented by Fogg (1998) and Acemoglu et al. (2024), we propose ScamGPT-J as a practical embodiment of LLMs as persuasive tools, with a specific focus on guiding human behavior in the context of scam prevention. Our paper adds to the literature on specific-use LLMs by offering an LLM as a persuasive technology for increasing individual resilience against scams.

## Our Proposed Solution: ScamGPT-J

In this section, we introduce our strategy aimed at implementing measures to bolster the resilience of individuals against becoming victims of scams. Our approach adds a proactive, non-intrusive layer of protection. We focus on the role of our custom-tuned generative LLM, ScamGPT-J, in countering instant messaging scams. We discuss the foundational principles that make this model an effective tool within the realm of persuasive technology, serving both as a safeguard and an educational resource.

## *Overview of our ScamGPT-J Solution*

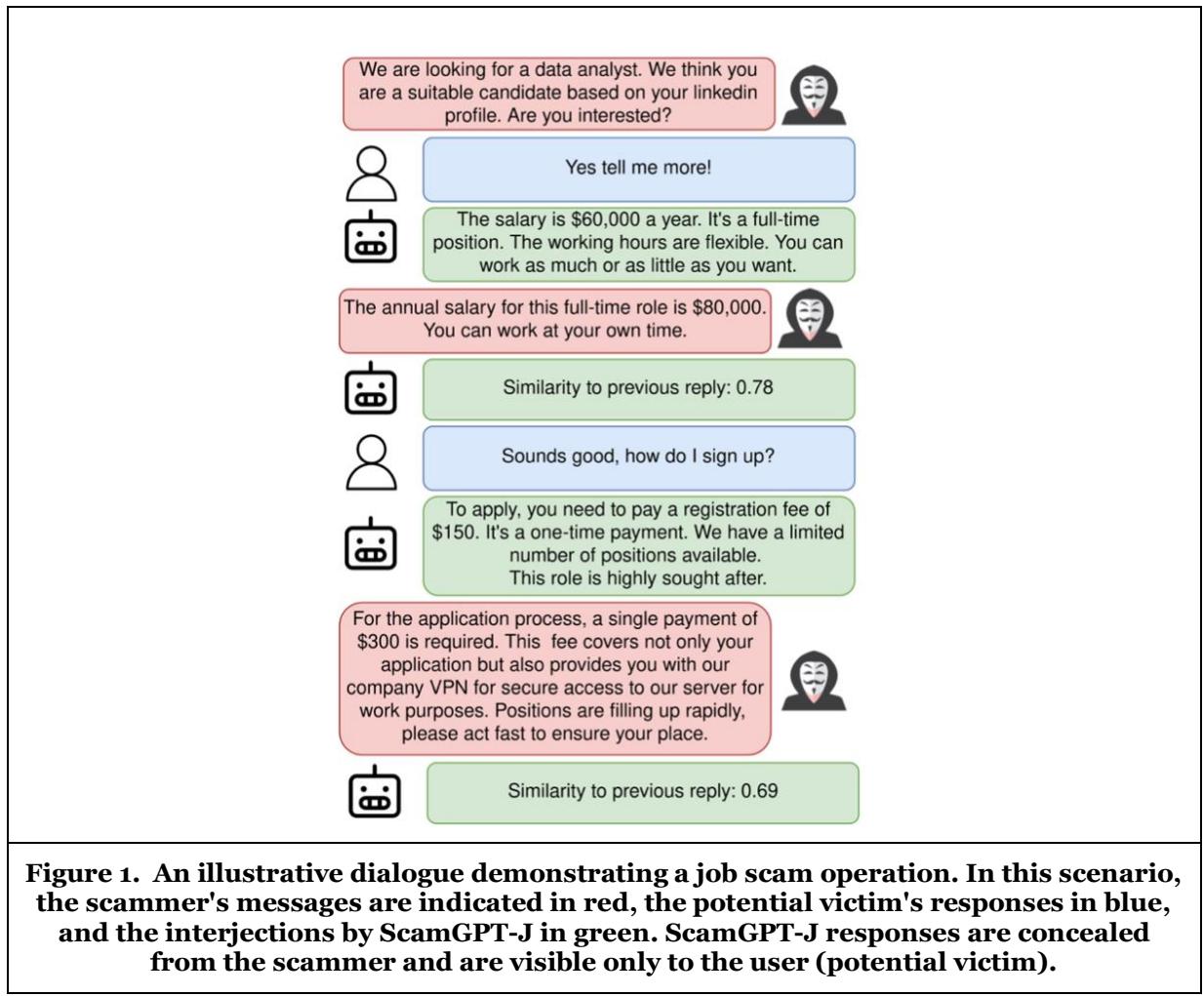

**Figure 1. An illustrative dialogue demonstrating a job scam operation. In this scenario, the scammer's messages are indicated in red, the potential victim's responses in blue, and the interjections by ScamGPT-J in green. ScamGPT-J responses are concealed from the scammer and are visible only to the user (potential victim).**

The risk of scams in digital communication necessitates robust, proactive defenses. Here, we present ScamGPT-J, an innovative AI solution built upon GPT-J (Wang & Komatsuzaki, 2021), an open-source LLM. ScamGPT-J is designed to act as a powerful shield against online fraud. It actively engages in potential scam conversations, generating real time responses that a scammer might use. The central idea is that ScamGPT-J, having been trained to mimic responses like a scammer, would only generate believable replies when the conversation is likely a scam. The user, fully aware that ScamGPT-J is trained to simulate a scammer's behavior, will carefully compare the responses from both the potential scammer and ScamGPT-J. This informed perspective enables the user to make a more accurate judgment about whether the conversation is likely to be a scam. Figure 1 graphically represents the complex interplay of communication among a scammer, a potential victim, and ScamGPT-J.

- The scammer begins the dialogue with an enticing job offer, attempting to lure the potential victim into a scam.
- The potential victim continues the dialogue by showing interest in the job offer.
- ScamGPT-J intervenes by generating responses that resemble those a scammer would make. These generated messages are based on the ongoing dialogue between the scammer and the potential victim.

- After the scammer responds to the victim, ScamGPT-J analyzes his text and computes a similarity score to quantify its resemblance to previous messages. This score is immediately reported to the potential victim.
- ScamGPT-J repeats this process after every reply from the user. All interactions from ScamGPT-J are confidentially shown only to the potential victim, keeping the scammer unaware of ScamGPT-J's presence and analysis.

## Hypothesis

We hypothesize that ScamGPT-J's predictive capabilities significantly enhance the persuasiveness of identifying deceptive conversations beyond the binary classifications of traditional machine learning algorithms. Traditional models, which label interactions merely as "scam" or "not a scam", may fail to capture the nuanced dynamics of scam tactics (Kawintiranon et al., 2022), often only providing victims with a generalized warning that lacks specific context. Furthermore, when confronted with information that conflicts with their cognitive beliefs, individuals tend to become defensive and reinforce their beliefs (Gillespie, 2020). Consequently, binary scam classification models, no matter how accurate they may be, are insufficient to convince potential victims to disengage with a scam.

In contrast, ScamGPT-J forecasts the scammer's likely responses, offering tangible, anticipatory evidence of a scam in progress. This forward-looking approach does more than signal potential danger; it aligns with the *Heuristic and Systematic Models of Information Processing* theory proposed in the field of social psychology (Chaiken, 1980). The theory suggests that individuals are more susceptible to heuristic persuasion, which requires less cognitive effort, and relies on non-content cues when they exhibit low involvement with the message subject. Non-content cues refer to peripheral factors, like emotional appeal or presentation style, which influence decision-making independently of the message's actual content. By providing specific, predicted dialogue from potential scammers, ScamGPT-J encourages individuals to engage more deeply with the conversation, which in turn catalyzes their shift to systematic processing, that focuses their attention more closely on the contents of the messages. Content cues pertain to the logical, factual, and argumentative quality of a message, including its factual data and relevance. These cues require detailed cognitive engagement, as they focus on critically analyzing the substance and validity of the information presented. When individuals engage in systematic processing, they focus on these content cues, analyzing and critically evaluating the message based on its merits and logical coherence. This contrasts with heuristic processing, which relies more on non-content cues. Under this paradigm of reasoning, individuals would be better able to analyze and detect any evidence of deception and realize for themselves if they are in the process of being scammed.

Additionally, we believe ScamGPT-J plays an important role in counteracting the anchoring effect (Furnham & Boo, 2011) that scammers may instill in their victims. The anchoring effect is a cognitive bias wherein individuals latch onto the initial and incomplete information presented to them (Tversky & Kahneman, 1974), forming an "anchor" that skews their perception of plausible realities. This effect is often exacerbated by *Behavioral Confirmation Bias* (Rosenthal & Jacobson, 1968), where individuals' expectations shape their perceptions and actions, leading them to interpret ambiguous information in a way that confirms their initial beliefs. As a result, individuals seek ways to validate what they believe to be the anchored reality, further cementing the effect (Chapman & Johnson, 1999; Mussweiler & Strack, 2001). In the context of instant messaging scams, the anchoring effect manifests as victims forming beliefs from early interactions that their messaging counterpart is genuine. This explains why victims often refuse to believe they are involved in a scam, even when presented with new evidence. ScamGPT-J mitigates this anchoring effect by employing a variation of the "consider-the-opposite" strategy proposed by Mussweiler et al. (2000). This strategy involves consistently presenting the individual with anchor-inconsistent knowledge, which is credible information that contradicts the initial beliefs that formed the anchor, thereby reducing their confidence in their initial assumptions. As individuals observe that ScamGPT-J, an AI tool designed to mimic scammers, consistently produces predictions that accurately reflect scammer behavior, they may gradually reconsider the genuineness of the person they are interacting with. Consequently, their defenses, which may have been lowered due to the anchoring effect, are raised, providing better protection against potential scams.

In essence, our hypothesis posits that ScamGPT-J's method of predictive modeling adds significant depth to scam analysis, transforming how potential victims perceive and react to deceptive tactics. It moves beyond mere detection, offering a persuasive, educative experience that empowers individuals to recognize and evade scams. This approach could have a profound psychological impact, shifting users from a state of passive awareness to an active defense against online fraud. Figure 2 provides a visual narrative of the interaction between a scammer, a potential victim, and our AI-based intervention tool, ScamGPT-J. It illustrates a four-step process beginning with a scammer initiating contact with a potential victim.

- The scammer builds a deceptive rapport with the victim.
- The scammer attempts to have the victim transfer money or divulge bank details.
- The victim, with the assistance of ScamGPT-J, observes that the scammer's replies share a resemblance to the responses predicted by the AI system.
- The victim recognizes the shared patterns in the scammer's and ScamGPT-J's responses and realizes the scam.

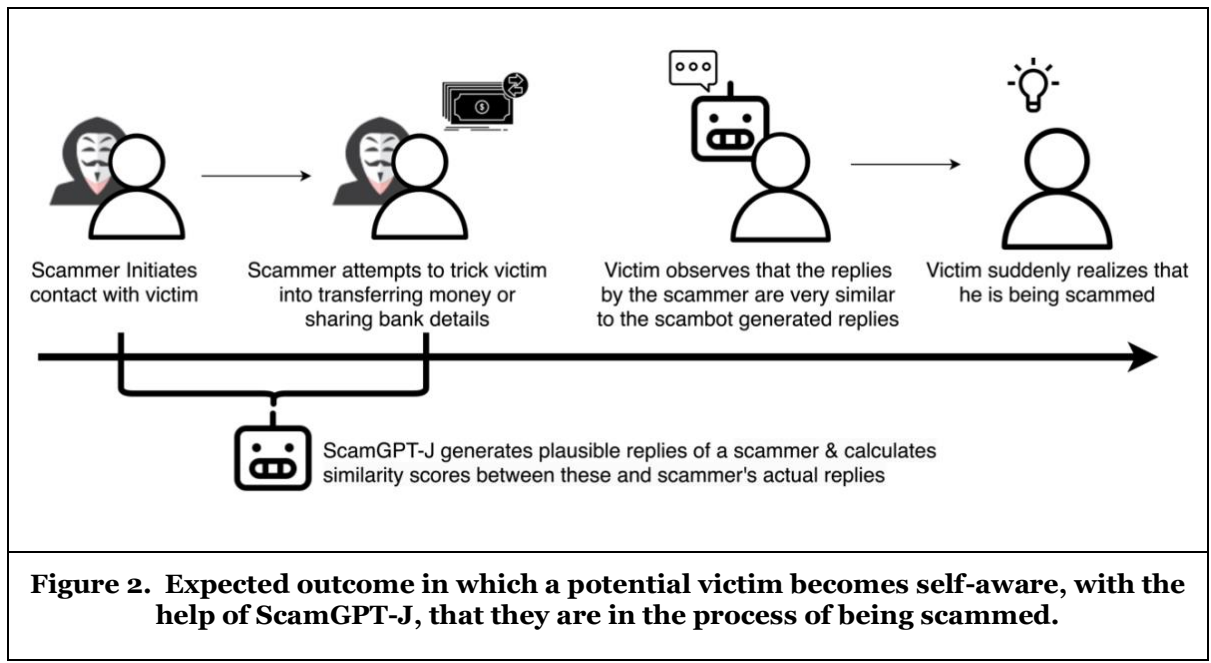

**Figure 2. Expected outcome in which a potential victim becomes self-aware, with the help of ScamGPT-J, that they are in the process of being scammed.**

## Model Architecture

Our model utilizes the open-source GPT-J model developed by EleutherAI, tailored through soft prompt tuning with the Hugging Face PEFT library. Selecting an open-source model as our base model is an intentional decision to ensure transparency, which is critical in applications requiring trust, reliability, and adaptability. The choice of an open-source model also ensures independence from the operational decisions of the company that developed the model. Thus, our model's functionality remains stable and reliable, unaffected by potential changes in EleutherAI's or any other organization's operations.

For soft prompt tuning, we use the PEFT library. This approach involves appending a trainable tensor to GPT-J's input embeddings, creating a soft prompt. We initiate this process with a specific seed prompt: "*Assuming you are a scammer, your goal is to trick a victim to give you money*". This seeds the model with a relevant starting context, from which it learns and optimizes the embeddings for the soft prompt. The dynamic nature of these prompts, learned through backpropagation and fine-tuned with diverse labeled examples, allows the model to adapt specifically to the subtleties of scammer communications. PEFT facilitates an efficient adaptation of GPT-J, minimizing parameter changes and avoiding the need for full model copies for each task.

We then conduct a comparative analysis of our model's effectiveness against the baseline performance of an untuned GPT-J. To achieve this, we utilize Sentence-BERT (Reimers & Gurevych, 2019) to compute the cosine similarity between responses generated by the LLM and the actual follow-up messages within our validation set of scam conversations. We hypothesize that ScamGPT-J demonstrates a higher cosine similarity score across these scam conversation scenarios (i.e., generate more convincing scammer-like replies based on previous interactions). The overall architecture is as shown in Figure 3.

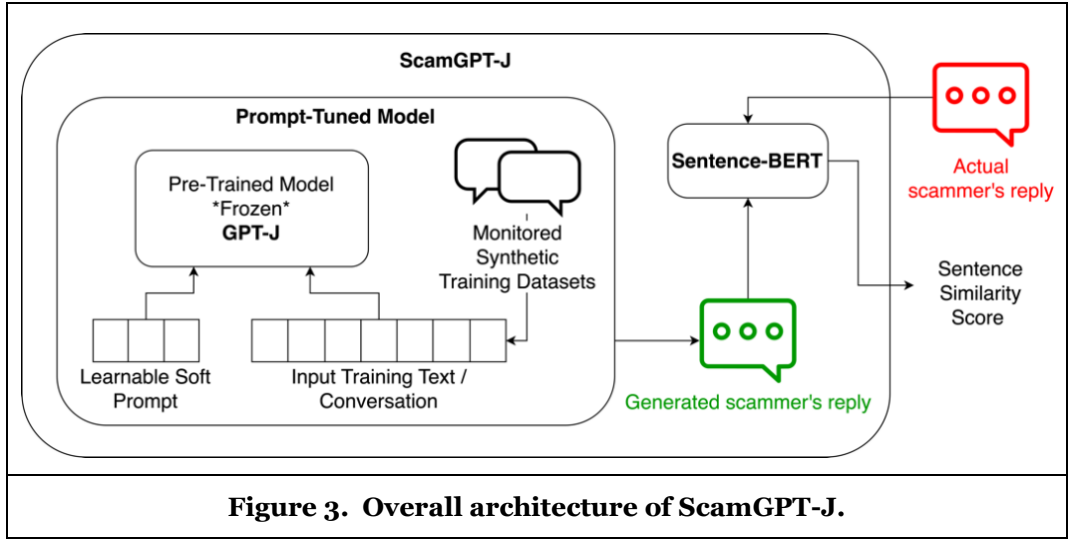

**Figure 3. Overall architecture of ScamGPT-J.**

## *Generating Prompt-Tuning Training Dataset*

We curate the training dataset for prompt tuning through an extensive and meticulous process, aimed at creating high-quality scam-related conversations. We begin by utilizing ChatGPT-4 to produce an initial set of 90 conversation samples. We obtained the sample outputs by feeding the platform with various inputs such as screenshots of actual scam message conversations found within news articles and police advisories and sophisticated hand-crafted scenarios based on actual conversations with scammers that we manually obtained. Our synthetic conversations revolve around four major scam categories: Authority, Job, Love, and Investment.

We then utilize OpenAI's LLM, GPT-3.5 Turbo, via its API to automate the expansion of our dataset. By using the initial 90 conversations as inputs into the API, we generated 10 new variants for each conversation, altering elements like names and conversation styles while retaining the core context. This process adds diversity to our dataset and ensures that the training data reflects the nuances of each scam type. However, given the occasionally inconsistent outputs from the OpenAI API, not all automatically generated conversations met our quality standards. An additional layer of quality control was conducted via manual vetting, where conversations that fell short of our criteria were identified and corrected manually.

Although we opted to generate the majority of the dataset, this decision does not preclude the manual collection of real scam messages. We collected the real scam messages by receiving text messages from scammers and prompting them to converse further to capture more of their conversation patterns and nuances. However, the quantity of real messages proved insufficient for effectively fine-tune a large language model. Additionally, the use of real messaging data raises significant privacy concerns, as these communications are typically confidential and private to individuals. To address both the data scarcity and privacy issues (Reiter, 2011), the collected real messages were utilized as seed data for generating additional variations (Endres et al., 2022). This approach allowed us to construct a comprehensive and diverse dataset that accurately represents the nuances of scam-related communications while mitigating the ethical and legal challenges associated with using real-world data.

At the end of this process, our dataset comprises a total of 902 conversations. We allocate 812 for training and reserve 90 for validation (i.e., to accurately assess the model's performance and generalization capabilities).

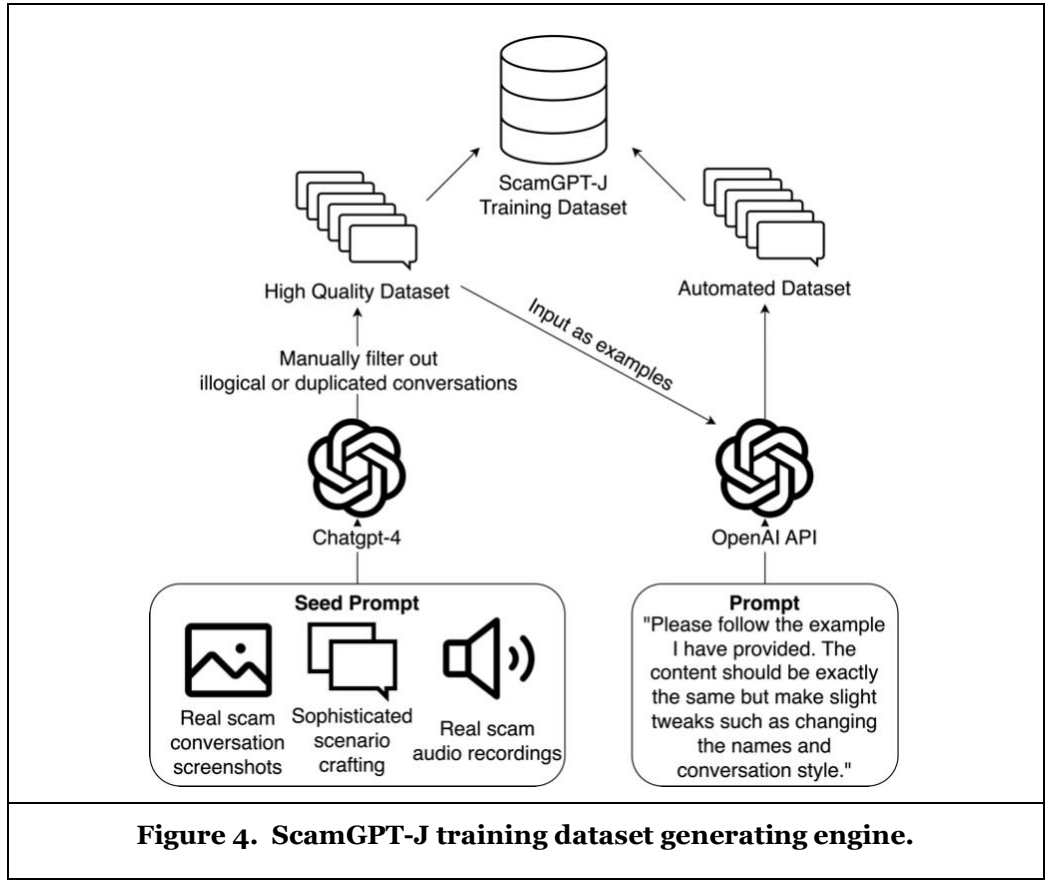

**Figure 4. ScamGPT-J training dataset generating engine.**

## Experiment

The assessment of the utility and efficacy of novel artifacts is an essential and integral process in the field HCI. This evaluation can adopt multiple methodologies, including analytical, case study, experimental, field study, and simulation approaches (Hevner et al., 2004). However, not all these methods are feasible due to resource constraints and ethical considerations.

Additionally, the evaluation process must account for societal impacts, both intended and unintended, of the artifacts (De Leoz & Petter, 2018). Given the sensitivity of the societal issues at hand, we have deliberately chosen to exclude field studies from our evaluation of ScamGPT-J. The ethical complexities of potentially influencing real scam cases necessitate this caution. We aim to avoid trialing our artifact in scenarios where it could inadvertently lead to negative outcomes, such as increasing a victim's susceptibility to a scam. Our evaluation methodology for ScamGPT-J involves a two-step process. Initially, we assess its efficacy against simulated conversations using well-defined metrics. This is followed by an experimental application with actual potential users, from whom we solicit feedback to understand its real-world behavioral impact. This approach is in accordance with the established compositional styles for technological artifact evaluation (Prat et al., 2015).

We believe that combining these two evaluation methods provides a comprehensive assessment of ScamGPT-J. This strategy ensures not only the technical robustness of our artifact but also its societal value.

### Technical Evaluation: ScamGPT-J vs GPT-J

In our technical evaluation, we assess the performance of our tuned LLM ScamGPT-J, against the original, untuned GPT-J model. This assessment is necessary to determine the efficacy of our tuning efforts in enhancing the model's ability to replicate scam conversations.

The evaluation centers around a dataset of 90 validation conversations, reflective of various scam scenarios. For each conversation, we simulate a response generation task. During the model's turn to reply, both ScamGPT-J and GPT-J are given the previous two interactions within the conversation as context to generate a response. This approach ensures that each model's output is based on a consistent and relevant context. Subsequently, we use Sentence-BERT (Reimers & Gurevych, 2019) to calculate the cosine similarities between the models' generated replies and the actual replies in the validation data. A value approaching one indicates a greater resemblance between the generated response and the actual subsequent reply, zero signifies orthogonality, and minus one denotes a response with a meaning opposite to that of the actual reply. Subsequently, we will determine the mean and max similarity scores for each validation conversation. As demonstrated in the conversation showcased in Figure 1, the *mean similarity score* is calculated as (0.78 + 0.69) / 2 = 0.74, while the *max similarity score* recorded is 0.78. After calculating the mean and max similarity scores for each of the 90 validation conversations, we calculate their overall averages and present these figures in the top two columns of Table 1. To further illustrate the superior capability of our ScamGPT-J model in generating scam-like responses, we detail the frequency with which ScamGPT-J outperforms GPT-J across these validation conversations indicated in the third column. Additionally, we have conducted a paired t-test on the scores from the 90 validation conversations, achieving a 99% confidence level that our prompt-tuned model, ScamGPT-J, demonstrates significant improvement over GPT-J.

|  | Mean similarity | Max similarity |
|---|---|---|
| ScamGPT-J | **0.433** | **0.622** |
| GPT-J | 0.329 | 0.525 |
| Instances of ScamGPT-J > GPT-J | 80 | 73 |
| Paired t-test across 90 validation conversations | | |
| p-value | **2.3e-10** | **3.6e-07** |
| t-statistic | 6.73 | 5.29 |

**Table 1. Overall performance statistics for ScamGPT-J and GPT-J. The first two columns show that ScamGPT-J can generate better scammer-like responses than GPT-J, as evidenced by the higher mean and max similarity scores. The third column reveals that ScamGPT-J outperformed GPT-J in 80 out of 90 validation conversations for mean similarity, and in 73 out of 90 validation conversations for max similarity. A paired t-test confirms with 99% confidence that the ScamGPT-J model significantly surpasses GPT-J in performance.**

### *User Experience and Feedback*

We create a survey that allows users to interact with ScamGPT-J and evaluate the quality and efficacy of its responses. The overall evaluation survey architecture is demonstrated in Figure 5.

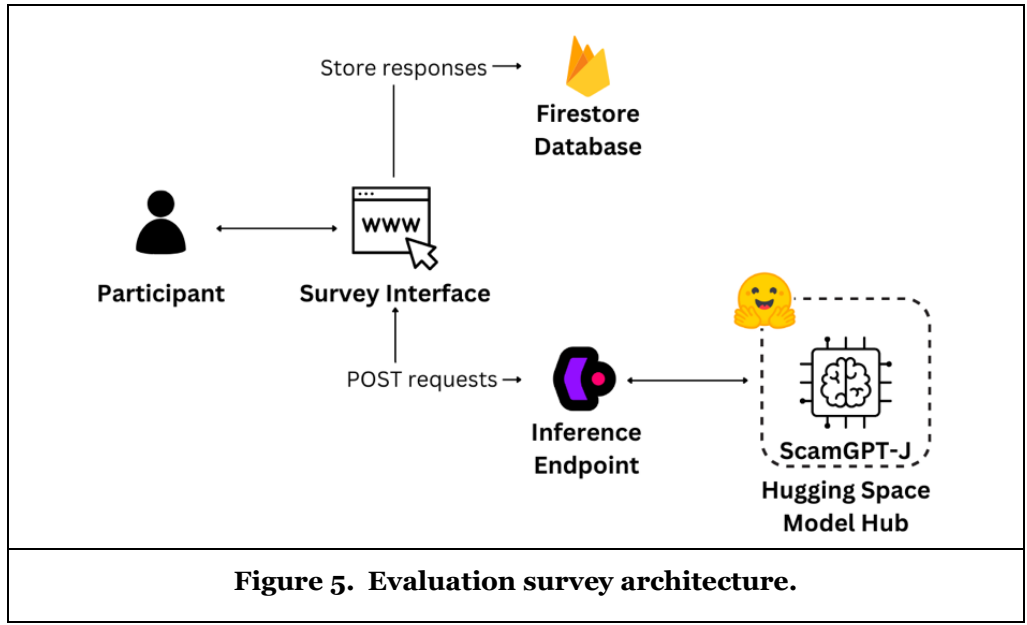

**Figure 5. Evaluation survey architecture.**

### Survey Platform Development and Deployment

We use a custom web application as a platform for volunteers to take part in our evaluation survey. The web application is connected to an API endpoint that allows survey participants to interact with our deployed LLM in real-time. Survey responses are collected in a Firestore Database for recording and processing.

### User Engagement and Response Collection

Our survey involves a group of 20 volunteers, each of whom has prior experience in scam conversations, a criterion that is essential to ensure relevant and informed feedback on ScamGPT-J. To manage participant data and maintain anonymity, we generate and distribute unique survey keys for each volunteer. These keys serve not only as distinct identifiers for the responses but also ensure that the collected data in our database was exclusively from these selected volunteers, without compromising their personal details.

We incorporate a double-blind methodology to ensure objectivity and fairness. Participants are randomly assigned either the ScamGPT-J model or an untuned GPT-J model for interaction, with no knowledge of which LLM they are interacting with. This approach is akin to a randomized controlled trial, where the users which are assigned the ScamGPT-J model are the treatment group.

The focus of our study is to observe the variation in responses to non-scam conversations. Given that GPT-J is a general-purpose LLM, we expect it to maintain consistency in responding to conversations regardless of their context. In contrast, ScamGPT-J, being specialized, is anticipated to demonstrate a deviation from standard responses in scam contexts. This aspect of our research is particularly important to understand the model's effectiveness in real-world scenarios, where the distinction between scam and legitimate interactions is a central function of our artifact.

The survey process commences with participants providing logs of three distinct conversations from their instant messaging exchanges. We require that these logs include at least one known scam conversation and one legitimate conversation, to ensure a comprehensive evaluation of the model's performance in varied contexts. The conversation logs are then processed through the LLM, with the responses displayed to the participants for evaluation. Participants, knowing the actual context of each conversation, are asked to assess whether the replies provided by the text generative model are believable and contextually appropriate. Additionally, they are required to identify the nature of each conversation, distinguishing between scam and legitimate interactions to aid our evaluation efforts.

In the final stage of the survey, we seek the participants' opinions on whether the model they interact with is effective in helping the participants discern potential scam scenarios by rating its usefulness on a scale from 1 to 5. The responses to this question allow us to assess ScamGPT-J on not only its technical ability but also its usefulness for scam mitigation purposes.

**Results and Discussion**

|  |  | GPT-J | ScamGPT-J |
|---|---|---|---|
| Scam Conversations | Context suited response | 3 | 14 |
|  | Non-context suited response | 16 | 2 |
| Normal Conversations | Context suited response | 9 | 4 |
|  | Non-context suited response | 2 | 10 |
| Total | | 30 | 30 |
| Average Usefulness Score (out of 5) | | 1.8 | **4.4** |

**Table 2. Survey results evaluating GPT-J vs. ScamGPT-J.**

The results of the evaluation survey are summarized in Table 2. Overall, the results are in line with our expectations. ScamGPT-J demonstrated a high degree of accuracy in mimicking scammer responses, successfully doing so in 14 out of 16 scam conversations. This starkly contrasts with GPT-J's performance, where it only correctly responded in 3 out of 19 scam scenarios. These findings unequivocally show that the specialized tuning of ScamGPT-J has made it adept at behaving like an actual scammer, far surpassing the capabilities of a general-purpose model like GPT-J. In terms of handling normal conversations, ScamGPT-J's performance was distinctively different. It produced context-inappropriate responses in 10 out of 14 non-scam dialogues, whereas GPT-J accurately responded in 9 out of 11 similar instances. This difference highlights ScamGPT-J's targeted effectiveness in scam scenarios, indicating its specialized design and functionality.

The survey results strongly support the idea that ScamGPT-J is an effective tool for identifying scam conversations. Its ability to closely mirror a scammer's responses in scam contexts and to diverge in normal conversations is a robust indicator for scam detection. This effectiveness is further backed by user evaluations of the models' ability to help discern if a conversation is a scam or not, with ScamGPT-J receiving a high average usefulness score of 4.4 out of 5. GPT-J's significantly lower score of 1.8 in the same context clearly illustrates the limitations of general-purpose LLMs in scam detection. These findings assert the critical role and superior performance of a specialized LLM like ScamGPT-J in identifying scams, thereby validating its design, and emphasizing its practical utility in combating online scams.

## Limitations and Future Work

In our research, ScamGPT-J was trained using synthetic conversations. While these conversations proved to be of adequate quality to elicit the desired behavior from the LLM, future iterations of this work could benefit from incorporating a corpus of real instant messaging scam conversations to enhance the model's robustness. Our study's user assessment faced constraints due to technical and time limitations. Participants interacted with our system in a restricted manner, and the conversations they provided as inputs were evaluated ex-post. Additionally, we recognize that our participants may have prior exposure to scam scenarios and may exhibit a bias when evaluating our artifact. A more extensive evaluation could be conducted in future research. Ideally, this would involve a larger group of participants who have no or limited exposure to scams with control and test groups evaluating their ability to detect scams with the support of ScamGPT-J versus without the support of LLMs. The implementation of such an experiment may involve the artifact being integrated directly into an instant messaging service, allowing participants to experience the system's responses in real-time as the conversation unfolds, potentially offering a more accurate assessment and mitigating the effects of look-back bias. We are currently working to secure the resources necessary for such a large-scale controlled experiment by collaborating with public agencies and financial institutions.

Furthermore, with our current framework, it is possible that there may be more false positives than false negatives upon implementation. In real-life contexts, messages from unknown phone numbers often originate from telemarketing, sales, or legitimate authorities, and their conversation style may closely resemble that of scam messages. However, the consequences of false positives (where users incorrectly identify a legitimate message as a scam) are generally less severe than false negatives (where users fail to recognize an actual scam). In the case of false positives, legitimate authorities can typically find alternative methods to reach the user if the message is urgent. On the other hand, false negatives can result in significant financial loss, and in many cases, the scammed funds are unrecoverable. Therefore, while the occurrence of false positives is a concern, the greater risk lies in false negatives, making it essential to prioritize minimizing them.

Another critical point to highlight is the risk of users becoming overly reliant on the tool, particularly if they focus solely on the similarity score provided by ScamGPT-J to determine whether a conversation is a scam. This overreliance could reduce critical engagement with the information presented, potentially leading users to accept the AI's judgment without sufficient scrutiny. The similarity score is intended as an additional signal or reference point to guide users in their decision-making process, rather than being the sole determinant. It is by no means a measurement of how scam-like the responses from the other party are, as the cosine similarity scores are context-free. Instead, they provide a reference for users to observe patterns, with no specific guidance on how to use the score. The similarity scores are not meant to be a recommendation metric but rather to indicate the general direction of the conversation. Users must interpret these scores themselves, and we acknowledge that there is subjectivity in how they choose to incorporate them into their decisions. For example, if a user notices consistently high similarity scores between the responses they receive and the ScamGPT-J simulated replies, it should signal that the person they are communicating with is more likely a scammer. However, how users account for or interpret these signals is a distinct HCI problem and is left for future work.

Our model is designed with a user-centric approach, aiming to empower users to make informed decisions and mitigate any scam-related anchoring effect (Furnham & Boo, 2011) using AI-generated insights to guide their own judgment. However, if users rely too heavily on the similarity score, they may fall into the trap of automation bias (Goddard et al., 2012), where they defer to the AI's output without critically assessing the context or content of the conversation. To mitigate this risk, future iterations of ScamGPT-J could incorporate features that encourage active user involvement such as offering prompts that guide users through a more systematic evaluation (Wash, 2020) of the conversation. By fostering a more interactive and reflective user experience, we can help ensure that the tool serves as a complement to human judgment, rather than a replacement, thereby promoting more balanced and accurate decision-making.

## Conclusion

This paper presents ScamGPT-J as a novel solution to combat scams by helping individuals realize scam situations through their own reasoning. We show how an LLM trained on a synthetic scam conversation corpus demonstrates a notable proficiency in predicting scammers' responses, while intentionally performing poorly in non-scam contexts. This dichotomy in response quality is central to our methodology: when ScamGPT-J aligns coherently with a given scam conversation, it suggests the presence of a scam operation. As demonstrated by our empirical results, ScamGPT-J can be a valuable and adaptable addition to the toolkits that law enforcement agencies use for combating scams. ScamGPT-J provides an effective intervention mechanism that can serve as a proactive means to assist at-risk individuals, guiding them in discerning and avoiding from scam ploys.

## Acknowledgements

This research / project is supported by the National Research Foundation, Singapore under its Industry Alignment Fund – Pre-positioning (IAF-PP) Funding Initiative. Any opinions, findings and conclusions or recommendations expressed in this material are those of the author(s) and do not reflect the views of National Research Foundation, Singapore.

## References

Abdel-Karim, B. M., Pfeuffer, N., Carl, K. V., & Hinz, O. 2023. "How AI-Based Systems Can Induce Reflections: The Case of AI-Augmented Diagnostic Work," *MIS Quarterly* (47:4), pp. 1395-1424.

Abraham, J., Junger, M., Koning, L., Njoki, C., & Rogers, S. 2023. *The Global State of Scams - 2023*.

Acemoglu, D., Makhdoumi, A., Malekian, A., & Ozdaglar, A. 2024. "When Big Data Enables Behavioral Manipulation," *American Economic Review: Insights*.

Almeida, T. A., Hidalgo, J. M. G., & Yamakami, A. 2011. "Contributions to the Study of SMS Spam Filtering: New Collection and Results," in: *Proceedings of the 11th ACM Symposium on Document Engineering*, pp. 259-262.

Aneke, J., Ardito, C., & Desolda, G. 2021. "Help the User Recognize a Phishing Scam: Design of Explanation Messages in Warning Interfaces for Phishing Attacks," in: *Proceedings of the 23rd International Conference on Human-Computer Interaction*, ACM, pp. 403-416.

Bada, M., Sasse, A. M., & Nurse, J. R. 2015. "Cyber Security Awareness Campaigns: Why Do They Fail to Change ?," in: *International Conference on Cyber Security for Sustainable Society*, pp. 118-131.

Canham, M., & Tuthill, J. 2022. "Planting a Poison SEAD: Using Social Engineering Active Defense (SEAD) to Counter Cybercriminals," in: *Proceedings of the 16th International Conference on Human-Computer Interaction*, ACM, pp. 48-57.

Chaiken, S. 1980. "Heuristic Versus Systematic Information Processing and the Use of Source Versus Cues in Persuasion," *Journal of Personality and Social Psychology* (39:5), pp. 752-766.

Chapman, G. B., & Johnson, E. J. 1999. "Anchoring, Activation, and the Construction of Values," *Organizational Behavior and Human Decision Processes* (79:2), pp. 115-153.

Chen, C., & Shu, K. 2024. Combating in the Age of LLMs: Opportunities and Challenges. *AI Magazine*.

Chen, T., & Kan, M.-Y. 2013. "Creating a Live, Public Short Message Service Corpus: The NUS SMS Corpus," *Language Resources and Evaluation* (47), pp. 299-335.

Cheung, T.-H., & Lam, K.-M. 2023. "FactLLaMA: Optimizing Instruction-Following Language Models with External Knowledge for Automated Fact-Checking," in: *Proceedings of the 2023 Asia Pacific Signal and Information Processing Association Annual Summit and Conference (APSIPA ASC)*, IEEE, pp. 846-853.

Cross, C. 2022. "Using Artificial Intelligence (AI) and Deepfakes to Deceive Victims: The Need to Rethink Current Romance Fraud Prevention Messaging," *Crime Prevention and Community Safety* (24:1), pp. 30-41.

De Leoz, G., & Petter, S. 2018. "Considering the Social Impacts of Artefacts in Information Systems Design Science Research," *European Journal of Information Systems* (27:2), pp. 154-170.

Dedeturk, B. K., & Akay, B. 2020. "Spam Filtering using a Logistic Regression Model Trained by an Artificial Bee Colony Algorithm," *Applied Soft Computing* (91), p. 106229.

Demszky, D., Yang, D., Yeager, D. S., Bryan, C. J., Clapper, M., Chandhok, S., Eichstaedt, J. C., Hecht, C., Jamieson, J., & Johnson, M. 2023. "Using Large Language Models in Psychology," *Nature Reviews Psychology* (2:11), pp. 688-701.

Derakhshan, A., Harris, I. G., & Behzadi, M. 2021. "Detecting Telephone-Based Social Engineering Attacks using Scam Signatures," in: *Proceedings of the 2021 ACM Workshop on Security and Privacy Analytics*, pp. 67-73.

Dove, M. 2020. *The Psychology of Fraud, Persuasion and Scam Techniques: Understanding What Makes Us Vulnerable*, Routledge.

Downs, J. S., Holbrook, M. B., & Cranor, L. F. 2006. "Decision Strategies and Susceptibility to Phishing," in: *Proceedings of the Second Symposium on Usable Privacy and Security*, pp. 79-90.

Edwards, M., Peersman, C., & Rashid, A. 2017. "Scamming the Scammers: Towards Automatic Detection of Persuasion in Advance Fee Frauds," in: *Proceedings of the 26th International Conference on World Wide Web Companion*, pp. 1291-1299.

Endres, M., Mannarapotta Venugopal, A., & Tran, T. S. 2022. "Synthetic Data Generation: A Comparative Study," in: *Proceedings of the 26th International Database Engineered Applications Symposium*, pp. 94-102.

Ezzeddine, F., Ayoub, O., Giordano, S., Nogara, G., Sbeity, I., Ferrara, E., & Luceri, L. 2023. "Exposing Influence Campaigns in the Age of LLMs: A Behavioral-Based AI Approach to Detecting State-Sponsored Trolls," *EPJ Data Science* (12:1), p. 46.

Fogg, B. J. 1998. "Persuasive Computers: Perspectives and Research Directions," in: *Proceedings of the SIGCHI conference on Human factors in computing systems*, ACM, pp. 225-232.

Furnham, A., & Boo, H. C. 2011. "A Literature Review of the Anchoring Effect," *The Journal of Socio-Economics* (40:1), pp. 35-42.

Gillespie, A. 2020. "Disruption, Self-Presentation, and Defensive Tactics at the Threshold of Learning," *Review of General Psychology* (24:4), pp. 382-396.

Goddard, K., Roudsari, A., & Wyatt, J. C. 2012. "Automation Bias: A Systematic Review of Frequency, Effect Mediators, and Mitigators," *Journal of the American Medical Informatics Association* (19:1), pp. 121-127.

Hevner, A. R., March, S. T., Park, J., & Ram, S. 2004. "Design Science in Information Systems Research," *MIS Quarterly*, pp. 75-105.

Jung, N., Wranke, C., Hamburger, K., & Knauff, M. 2014. "How Emotions Affect Logical Reasoning: Evidence From Experiments With Mood-Manipulated Participants, Spider phobics, and People with Exam Anxiety," *Frontiers in Psychology* (5), p. 570.

Kassem, R. 2023. "How Fraud Impacts Individuals' Wellbeing—Academic Insights and Gaps," *Journal of Financial Crime*.

Kawintiranon, K., Singh, L., & Budak, C. 2022. "Traditional and Context-Specific Spam Detection in Low Resource Settings," *Machine Learning* (111:7), pp. 2515-2536.

Kumar, S., & Gupta, S. 2024. "Legitimate and Spam SMS Classification Employing Novel Ensemble Feature Selection Algorithm," *Multimedia Tools and Applications* (83:7), pp. 19897-19927.

Ma, K. W. F., & McKinnon, T. 2021. "COVID-19 and Cyber Fraud: Emerging Threats During the Pandemic," *Journal of Financial Crime* (29:2), pp. 433-446.

Martin, J., Dubé, C., & Coovert, M. D. 2018. "Signal Detection Theory (SDT) is Effective for Modeling User Behavior Toward Phishing and Spear-Phishing Attacks," *Human Factors* (60:8), pp. 1179-1191.

Mishra, S., & Soni, D. 2023. "DSmishSMS-A System to Detect Smishing SMS," *Neural Computing and Applications* (35:7), pp. 4975-4992.

Mussweiler, T., & Strack, F. 2001. "The Semantics of Anchoring," *Organizational Behavior and Human Decision Processes* (86:2), pp. 234-255.

Mussweiler, T., Strack, F., & Pfeiffer, T. 2000. "Overcoming the Inevitable Anchoring Effect: Considering the Opposite Compensates for Selective Accessibility," *Personality and Social Psychology Bulletin* (26:9), pp. 1142-1150.

Prat, N., Comyn-Wattiau, I., & Akoka, J. 2015. "A Taxonomy of Evaluation Methods for Information Systems Artifacts," *Journal of Management Information Systems* (32:3), pp. 229-267.

Reimers, N., & Gurevych, I. 2019. "Sentence-BERT: Sentence Embeddings using Siamese BERT-Networks," in: *Proceedings of the 2019 Conference on Empirical Methods in Natural Language Processing and the 9th International Joint Conference on Natural Language Processing (EMNLP-IJCNLP)*, pp. 3982-3992.

Reinheimer, B., Aldag, L., Mayer, P., Mossano, M., Duezguen, R., Lofthouse, B., Von Landesberger, T., & Volkamer, M. 2020. "An Investigation of Phishing Awareness and Education Over Time: When and How to Best Remind Users," in: *Proceedings of the Sixteenth Symposium on Usable Privacy and Security (SOUPS 2020)*, pp. 259-284.

Reiter, J. P. 2011. "Data Confidentiality," *Wiley Interdisciplinary Reviews: Computational Statistics* (3:5), pp. 450-456.

Rosenthal, R., & Jacobson, L. 1968. "Pygmalion in the Classroom," *The Urban Review* (3:1), pp. 16-20.

Saidani, N., Adi, K., & Allili, M. S. 2020. "A Semantic-Based Classification Approach for an Enhanced Spam Detection," *Computers & Security* (94), p. 101716.

Shafi'I, M. A., Abd Latiff, M. S., Chiroma, H., Osho, O., Abdul-Salaam, G., Abubakar, A. I., & Herawan, T. 2017. "A Review on Mobile SMS Spam Filtering Techniques," *IEEE Access* (5), pp. 15650-15666.

Shillair, R., Esteve-González, P., Dutton, W. H., Creese, S., Nagyfejeo, E., & von Solms, B. 2022. "Cybersecurity Education, Awareness Raising, and Training Initiatives: National Level Evidence-Based Results, Challenges, and Promise," *Computers & Security* (119), p. 102756.

Tabone, W., & de Winter, J. 2023. "Using ChatGPT for Human–Computer Interaction Research: A Primer," *Royal Society Open Science* (10:9), p. 231053.

Tan, C., & Devaraj, S. 2023. 'I'll make sure you lose your job,' woman tells bank staff who tried to stop her from sending money to 'husband'. *The Straits Times*.

Teodorescu, M. H., Morse, L., Awwad, Y., & Kane, G. C. 2021. "Failures of Fairness in Automation Require a Deeper Understanding of Human-ML Augmentation," *MIS Quarterly* (45:3), pp. 1483-1500.

Tversky, A., & Kahneman, D. 1974. "Judgment under Uncertainty: Heuristics and Biases: Biases in judgments reveal some heuristics of thinking under uncertainty," *Science* (185:4157), pp. 1124-1131.

Van Dijk, E., & Zeelenberg, M. 2003. "The Discounting of Ambiguous Information in Economic Decision Making," *Journal of Behavioral Decision Making* (16:5), pp. 341-352.

Wang, B., & Komatsuzaki, A. 2021. *GPT-J-6B: A 6 billion parameter autoregressive language model*.

Wash, R. 2020. "How Experts Detect Phishing Scam Emails," *Proceedings of the ACM on Human-Computer Interaction* (4:CSCW2), pp. 1-28.

Whitty, M. T., & Buchanan, T. 2016. "The Online Dating Romance Scam: The Psychological Impact on Victims–Both Financial and Non-Financial," *Criminology & Criminal Justice* (16:2), pp. 176-194.

Wijaya, E., Noveliora, G., Utami, K. D., & Nabiilah, G. Z. 2023. "Spam Detection in Short Message Service (SMS) Using Naïve Bayes, SVM, LSTM, and CNN," in: *Proceedings of the 2023 10th International Conference on Information Technology, Computer, and Electrical Engineering (ICITACEE)*, pp. 431-436.

Yang, K.-C., & Menczer, F. 2024. "Anatomy of an AI-Powered Malicious Social Botnet," *Journal of Quantitative Description: Digital Media* (4), pp. 1-36.

Zhang, X., & Gao, W. 2023, November. "Towards LLM-Based Fact Verification on News Claims with a Hierarchical Step-by-Step Prompting Method," in: *Proceedings of the 13th International Joint Conference on Natural Language Processing and the 3rd Conference of the Asia-Pacific Chapter of the Association for Computational Linguistics (Volume 1: Long Papers)*, pp. 996-1011.

Zhang, Y., Wu, Q., Zhang, T., & Yang, L. 2022. "Vulnerability and Fraud: Evidence from the COVID-19 Pandemic," *Humanities and Social Sciences Communications* (9:1), pp. 1-12.

Zhuo, S., Biddle, R., Koh, Y. S., Lottridge, D., & Russello, G. 2023. "SoK: Human-Centered Phishing Susceptibility," *ACM Transactions on Privacy and Security* (26:3), pp. 1-27.